\documentclass[12pt]{iopart}

\newcommand{\bea}{\begin{eqnarray}}
\newcommand{\eea}{\end{eqnarray}}
\usepackage{graphicx}
\usepackage{epsfig}  
\begin{document}

\title[Correlations and fluctuations from lattice QCD]{Correlations and fluctuations from lattice QCD}

\author{Szabolcs Borsanyi$^1$, Zoltan Fodor$^{1,2,3}$, Sandor Katz$^2$, Stefan Krieg$^{1,3}$, Claudia Ratti$^{4,5}$ and Kalman Szabo$^1$}

\address{$^1$  
Dep. of Physics, Wuppertal University, Gaussstr. 20, D-42119 Wuppertal, Germany}
\address{$^2$ Inst. for Theoretical Physics, E\"otv\"os University, P\'azm\'any 1, H-1117 Budapest, Hungary} 
\address{$^3$ J\"ulich Supercomputing Centre, Forschungszentrum J\"ulich, D-52425
J\"ulich, Germany}
\address{$^4$ Dip. di Fisica Teorica, Universit\`a di Torino, via Giuria 1, I-10125 Torino, Italy}
\address{$^5$ INFN, Sezione di Torino}
\begin{abstract}
We present the new results of the Wuppertal-Budapest lattice QCD collaboration on flavor diagonal and non-diagonal quark number susceptibilities with 2+1 staggered quark flavors, in a temperature regime between 120 and 400 MeV. A Symanzik improved gauge and a stout-link improved staggered fermion action is utilized; the light and strange quark masses are set to their physical values. Lattices with $N_t=6,~8,~10,~12$ are used. We perform a continuum extrapolation of those observables for which the scaling regime is reached, and discretization errors are under control.
\end{abstract}

\maketitle

\section{Introduction}
\vspace{-.3cm}
A transition occurs in strongly
interacting matter from a hadronic, confined system at small temperatures and
densities to a phase dominated by colored degrees of freedom at large
temperatures or densities. This field of physics is particularly appealing because
the deconfined phase of QCD can be produced in the laboratory, in the
ultrarelativistic heavy ion collision experiments at CERN SPS, RHIC at
Brookhaven National Laboratory, ALICE at the LHC and the future FAIR at the
GSI. A lot of effort is invested, both theoretically and experimentally, in order to find observables which would unambiguously signal the QCD phase transition. 
Correlations and fluctuations of conserved charges have been proposed long ago to this purpose
\cite{Jeon:2000wg,Asakawa:2000wh}. The idea is that these quantum numbers have a very different value in a confined and deconfined system, and measuring them in the laboratory would allow to distinguish between the two phases.

Fluctuations of conserved charges can be obtained as linear combinations of diagonal and non-diagonal quark number susceptibilities, which can be calculated on the lattice at zero chemical potential \cite{hotQCD2,Borsanyi:2010bp}. In the present contribution we show the results of our collaboration on some of these observables, with 2+1 staggered quark flavors, in a temperature regime between 120 and 400 MeV.  The light and strange quark masses are set to their physical values. Lattices with $N_t=6,~8,~10,~12$ are used. Continuum extrapolations are performed for those observables for which discretization errors are under control and the scaling regime is reached.

\begin{figure}
\begin{minipage}{.48\textwidth}
\parbox{6cm}{
\scalebox{.61}{
\includegraphics{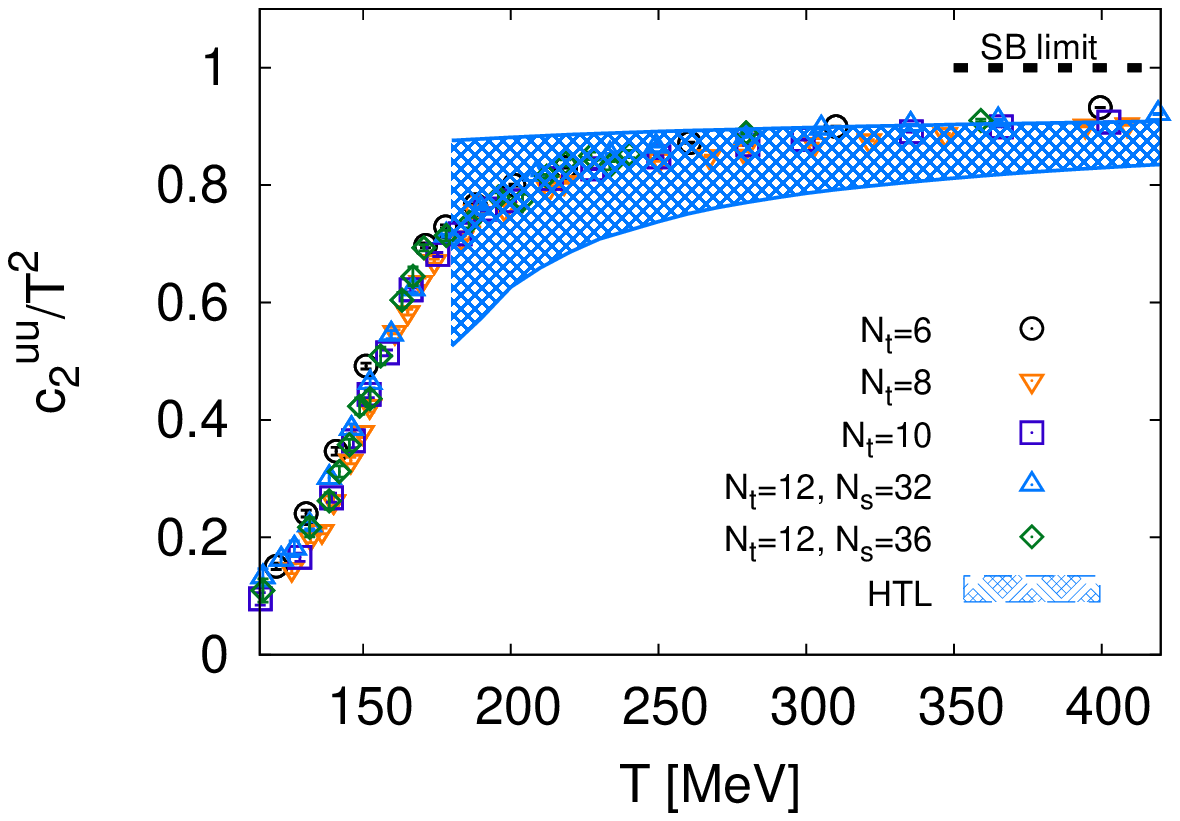}\\}}
\end{minipage}
\hspace{.1cm}
\begin{minipage}{.48\textwidth}
\parbox{6cm}{
\scalebox{.61}{
\includegraphics{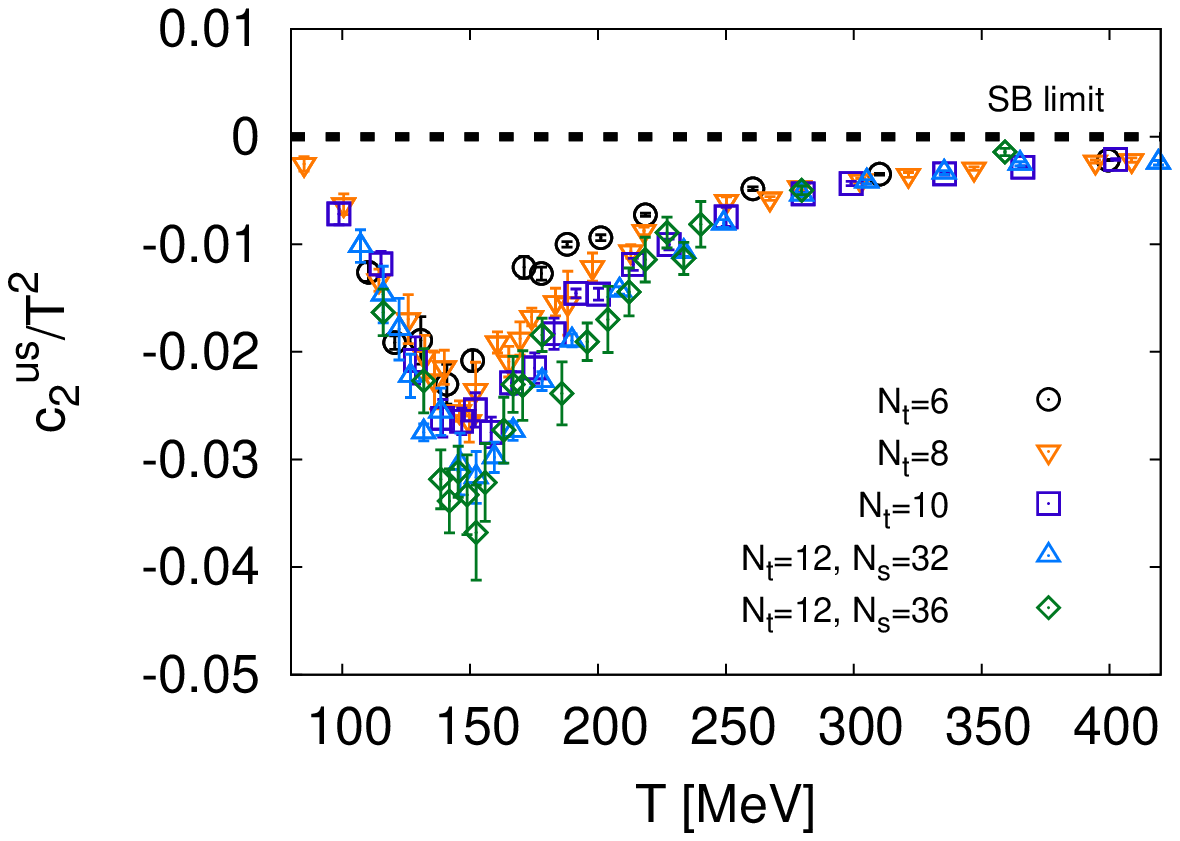}\\}}
\end{minipage}
\caption{
Left: Diagonal light quark susceptibility as a function of the 
temperature. Right: Non-diagonal $u-s$ susceptibility as a function of the temperature. In both panels, data are obtained on $N_t = 6,~ 8,~ 10$ and 12 lattices. Two different spatial volumes are considered for the latter: $32^3$ and $36^3$. The band in the left panel is the comparison with the Hard Thermal Loop predictions taken from Ref. \cite{Blaizot:2001vr}}
\label{fig3}
\vspace{-.4cm}
\end{figure}
\begin{figure}
\begin{minipage}{.48\textwidth}
\parbox{6cm}{
\scalebox{.61}{
\includegraphics{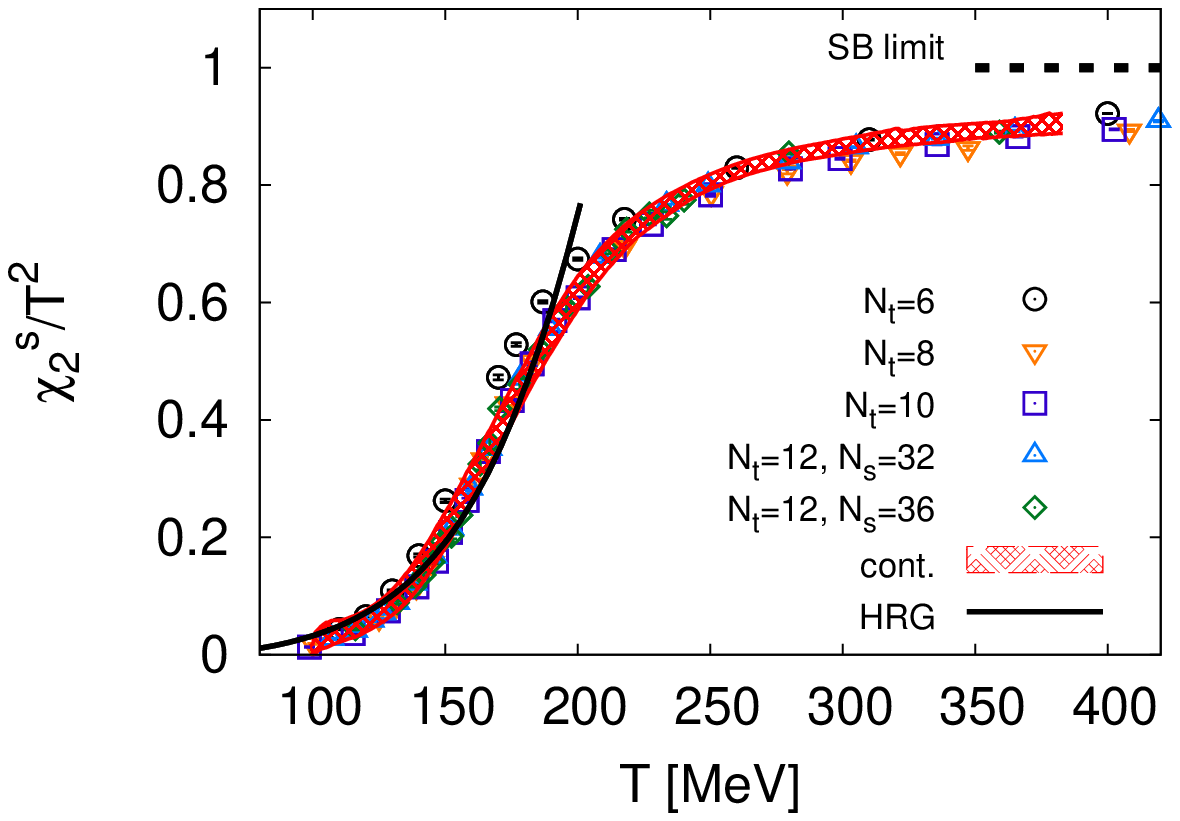}\\}}
\end{minipage}
\hspace{.6cm}
\begin{minipage}{.48\textwidth}
\parbox{6cm}{
\scalebox{.55}{
\includegraphics{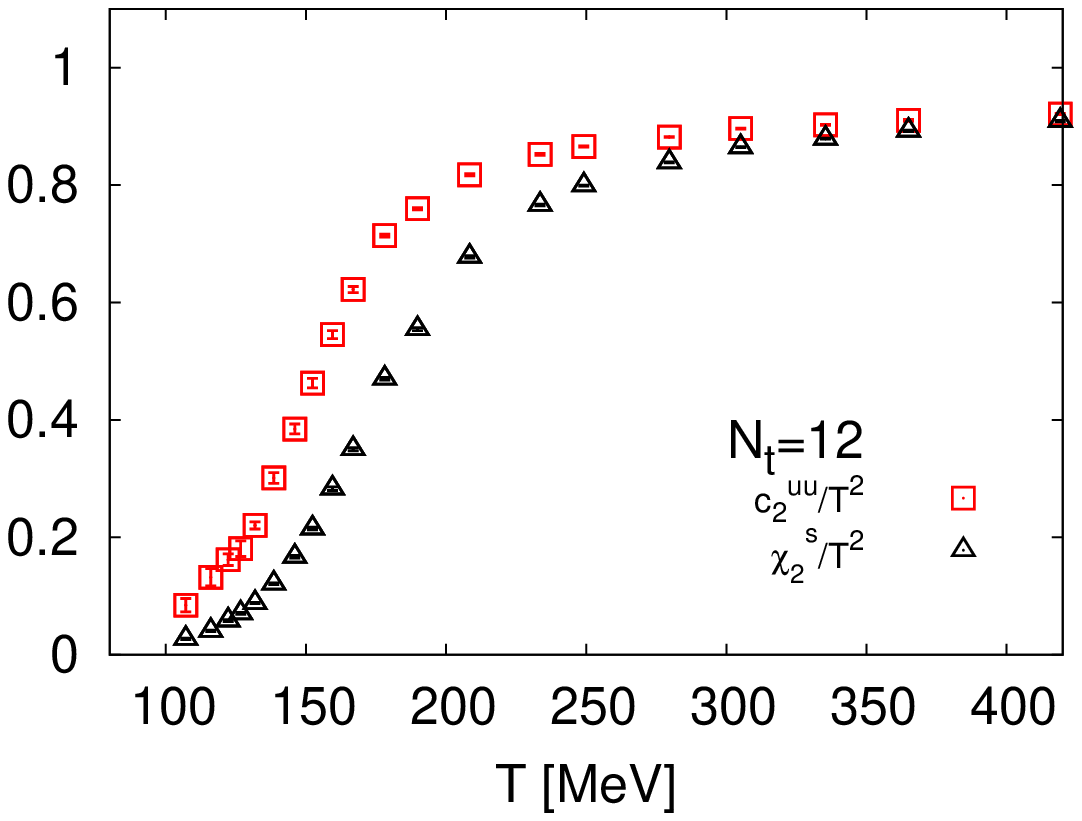}\\}}
\end{minipage}
\caption{
Left: Diagonal strange quark susceptibility as a function of the 
temperature. Data are obtained on $N_t = 6,~ 8,~ 10$ and 12 lattices. Two different spatial volumes are considered for the latter: $32^3$ and $36^3$. The red band indicates the continuum extrapolation. The black curve is the Hadron Resonance Gas (HRG) model prediction. Right: comparison between the light and strange quark susceptibility as functions of the temperature, for $N_t=12$.}
\label{fig4}
\vspace{-.5cm}
\end{figure}
\vspace{-.3cm}
\section{Results}
\vspace{-.3cm}
Quark number susceptibilities are defined as derivatives of the QCD pressure with respect to the quark chemical potentials. In particular, we consider the following ones:
\bea
\!\!\!\!\!\!\!\!\!\!\!\!\!\!\!\!\!\!\!\!\!\!\!\!\!\!\!\!\!\!\!\!\!\!
{c_{2}^{uu}=\left.\frac{T}{V}\frac{\partial^2\ln Z}{\partial\mu_{u}\partial\mu_{u}}\right|_{\mu_i=0}},~~~~~~
{c_{2}^{us}=c_{2}^{ds}=\left.\frac{T}{V}\frac{\partial^2\ln Z}{\partial\mu_{u}\partial\mu_{s}}\right|_{\mu_i=0}},
~~~~~~
{\chi_{2}^{s}=\left.\frac{T}{V}\frac{\partial^2\ln Z}{\partial\mu_{s}\partial\mu_{s}}\right|_{\mu_i=0}}.
\eea
The lattice action is the same as we used in \cite{6, 7}, namely a tree-level Symanzik improved gauge, and a stout-improved staggered fermionic action (see Ref. \cite{Aoki:2005vt} for details).  The stout-smearing 
\cite{Morningstar:2003gk} reduces the taste violation: this kind of smearing has one of the 
smallest taste violation among the ones used so far in the literature, together with the
hisq action recently proposed by the hotQCD collaboration~\cite{20,Bazavov:2010sb}.
In the left panel of Fig.~\ref{fig3} we show the diagonal light quark susceptibility as a function of the temperature. Results are obtained on five different lattices: $N_t=6,~8,~10$ and two different spatial volumes for $N_t=12$: $32^3$ and $36^3$. This observable reaches approximately 90\% of the Stefan-Boltzmann limit already at $T\sim400$ MeV. Notice that we do not perform a continuum extrapolation: this observable is pion-dominated at small temperatures and we need either finer lattices or an action which reduces even more the taste violation, in order to reach the scaling regime and give the correct continuum estimate. In the right panel of Fig.~\ref{fig3} we show the non-diagonal $u-s$ susceptibility, which measures the correlation between these different flavors. As we can see, this observable has a peak in the vicinity of the phase transition, and tends to zero for infinitely large temperatures. This observable is very noisy and before performing a continuum extrapolation we need to reduce the errorbars for the $N_t=12$ sets of data.
In the left panel of Fig.~\ref{fig4} we show the strange quark number susceptibility as a function of the temperature. For this observable we perform a continuum extrapolation, as well as a comparison to the HRG model predictions.
In the right panel of Fig. \ref{fig4} we show, for $N_t=12$, a comparison between light and strange susceptibilities. We notice that the rapid rise as a function of the temperature occurs at larger temperatures for strange quarks than for light quarks.

\begin{figure}
\begin{minipage}{.48\textwidth}
\parbox{6cm}{
\scalebox{.61}{
\includegraphics{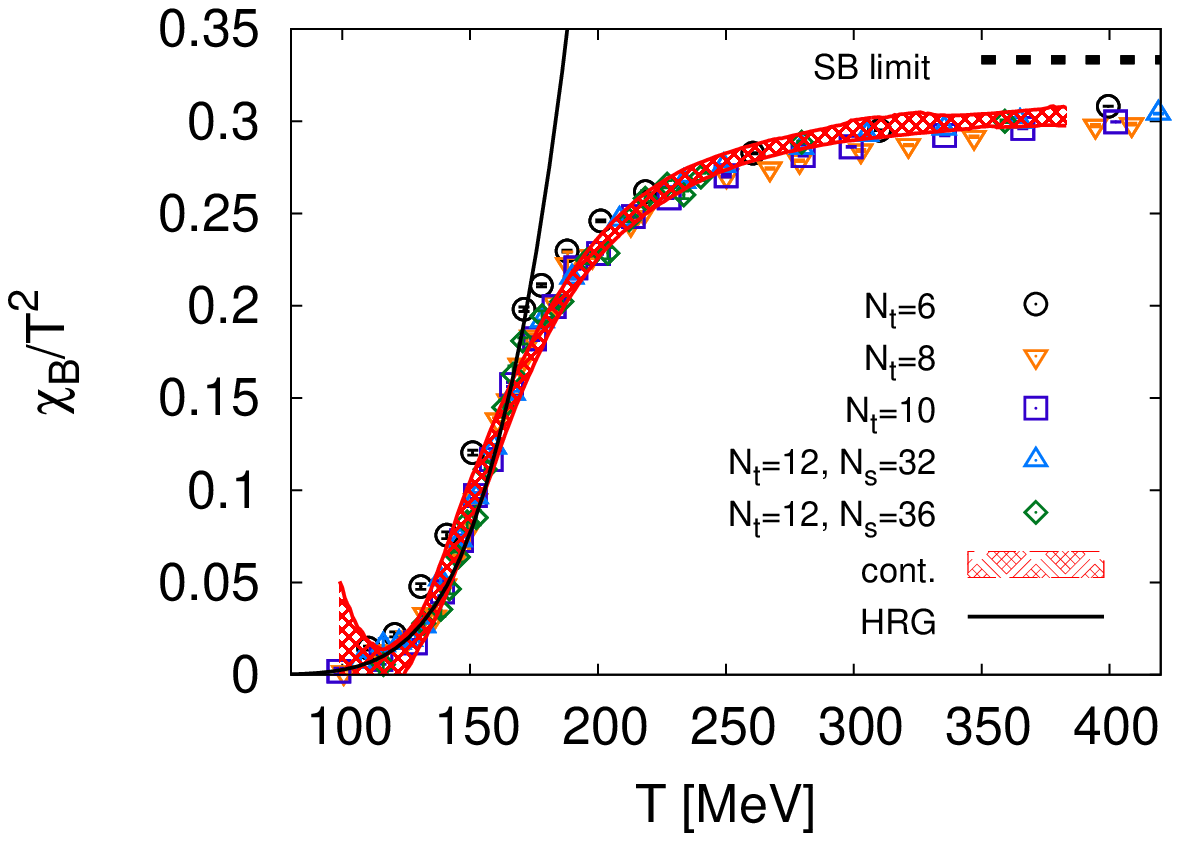}\\}}
\end{minipage}
\hspace{.1cm}
\begin{minipage}{.42\textwidth}
\parbox{6cm}{
\scalebox{.61}{
\includegraphics{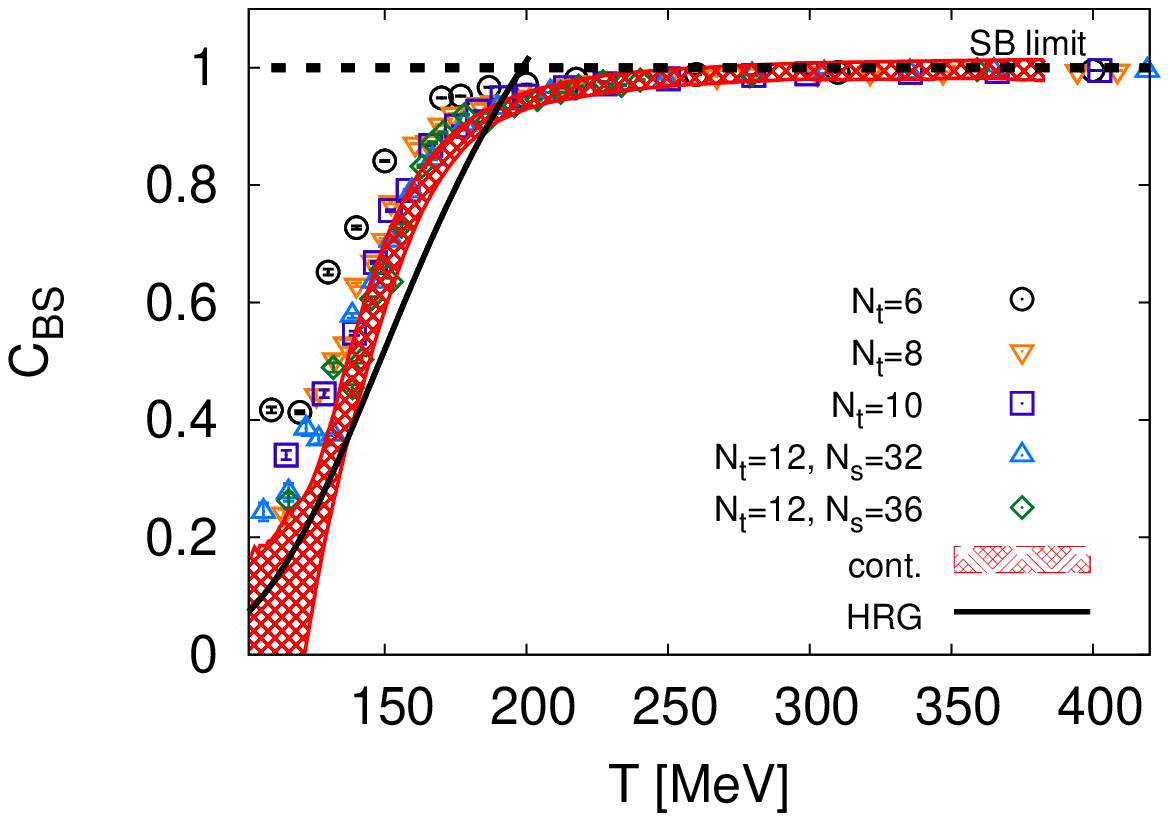}}}\\
\end{minipage}
\caption{
Left: Baryon number susceptibility as a function of the temperature. Right: Baryon-strangeness correlator as a function of the temperature. In both panels, results for $N_t=6,~8,~10$ and 12 lattices are shown. Two different spatial volumes are considered for the latter: $32^3$ and $36^3$. The red band is the continuum extrapolated result, the black curve is the prediction of the HRG model.}
\label{curv}
\vspace{-.5cm}
\end{figure}

Baryon number susceptibility is defined as a linear combination of quark number susceptibilities in the following way:
\bea
\!\!\!\!\!\!\!\!\!\!\!\!\chi_B=\frac19\left(2c_{2}^{uu}+\chi_{2}^{s}+2c_{2}^{ud}+4c_{2}^{us}\right),~~~~~~
\mathrm{with}~~~~~~{c_{2}^{ud}=\left.\frac{T}{V}\frac{\partial^2\ln Z}{\partial\mu_{u}\partial\mu_{d}}\right|_{\mu_i=0}}.
\eea
We show our results for this observable in the left panel of Fig.~\ref{curv}: a continuum extrapolation is performed, and compared to the HRG model results. The agreement is very good.
In the right panel of Fig. \ref{curv} we show the baryon-strangeness correlator that was proposed in Ref. \cite{Koch:2005vg} as a diagnostic of strongly interacting matter. It is defined as:
\bea
C_{BS}=1+\frac{c_{2}^{us}+c_{2}^{ds}}{\chi_{2}^{s}}.
\vspace{-.3cm}
\eea
The continuum extrapolation is in good agreement with the HRG model prediction for temperatures smaller than $T_c$.
\vspace{-.4cm}
\section*{Acknowledgements}
\vspace{-.2cm}
Work supported in part by the EU grant  (FP7/2007-2013)/ERC no. 208740. The work of C. R. is supported by funds provided by the Italian 
Ministry of Education, Universities and Research under the Firb Research Grant 
RBFR0814TT. 


\vspace{-.5cm}
\section*{References}
\vspace{-.4cm}
\begin{thebibliography}{10}

\bibitem{Jeon:2000wg}
  S.~Jeon and V.~Koch,
  Phys.\ Rev.\ Lett.\  {\bf 85}, 2076 (2000)

\bibitem{Asakawa:2000wh}
  M.~Asakawa, U.~W.~Heinz and B.~Muller,
  Phys.\ Rev.\ Lett.\  {\bf 85}, 2072 (2000)
\bibitem{hotQCD2}
A.~Bazavov {\it et al.},
  Phys.\ Rev.\  D {\bf 80}, 014504 (2009); S. Mukherjee, [arXiv:1107.0765 [nucl-th]].
\bibitem{Borsanyi:2010bp}
  S.~Borsanyi, Z.~Fodor, C.~Hoelbling, S.~Katz, S.~Krieg, C.~Ratti, K.~Szabo,             
  JHEP {\bf 1009}, 073 (2010)
\bibitem{6}
Y.~Aoki, Z.~Fodor, S.~D.~Katz and K.~K.~Szabo,
  Phys.\ Lett.\  B {\bf 643}, 46 (2006)
  
\bibitem{7}
Y.~Aoki, S.~Borsanyi, S.~Durr, Z.~Fodor, S.~Katz, S.~Krieg and K.~Szabo,
  JHEP {\bf 0906}, 088 (2009)
\bibitem{Aoki:2005vt}
  Y.~Aoki, Z.~Fodor, S.~D.~Katz and K.~K.~Szabo,
  JHEP {\bf 0601}, 089 (2006)
  
\bibitem{Morningstar:2003gk}
C.~Morningstar and M.~J.~Peardon,
  Phys.\ Rev.\  D {\bf 69}, 054501 (2004)
  
  \bibitem{20}
 A.~Bazavov and P.~Petreczky,
  PoS {\bf LAT2009}, 163 (2009)
  
\bibitem{Bazavov:2010sb}
  A.~Bazavov and P.~Petreczky,
  arXiv:1005.1131 [hep-lat].
 
 \bibitem{Blaizot:2001vr}
  J.~P.~Blaizot, E.~Iancu, A.~Rebhan,
  Phys.\ Lett.\  {\bf B523}, 143-150 (2001).
  [hep-ph/0110369].
 
 \bibitem{Koch:2005vg}
  V.~Koch, A.~Majumder and J.~Randrup,
  Phys.\ Rev.\ Lett.\  {\bf 95}, 182301 (2005)
 
   \end{thebibliography}
\end{document}